\documentclass{pnastwo}

\usepackage[dvips]{graphicx}

\usepackage{amssymb,amsfonts,amsmath}

\def\lapp{\ifmmode\stackrel{<}{_{\sim}}\else$\stackrel{<}{_{\sim}}$\fi}
\def\gapp{\ifmmode\stackrel{>}{_{\sim}}\else$\stackrel{>}{_{\sim}}$\fi}

\contributor{Published in Proceedings
of the National Academy of Sciences of the United States of America}
\url{www.pnas.org/cgi/doi/10.1073/pnas.0709640104}
\copyrightyear{2008}
\issuedate{Issue Date}
\volume{Volume}
\issuenumber{Issue Number}

\begin{document}



\title{Grand Unification of Neutron Stars}





\author{Victoria M. Kaspi\affil{1}{Department of Physics, McGill University, Montreal, Canada}}

\contributor{Submitted to Proceedings of the National Academy of Sciences
of the United States of America}

\maketitle

\begin{article}

\begin{abstract} 
The last decade has shown us that the observational properties of neutron
stars are remarkably
diverse.  From magnetars to rotating radio transients, from radio pulsars to 
`isolated neutron stars,' from central compact objects to millisecond pulsars, 
observational manifestations
of neutron stars are surprisingly varied, with most properties totally
unpredicted.  The challenge is to establish an overarching physical theory
of neutron stars and their birth properties that can explain this
great diversity.  Here I survey the disparate neutron stars classes, 
describe their properties, and highlight results made possible by
the {\it Chandra X-ray Observatory}, in celebration of its tenth anniversary.
Finally, I describe the current status of efforts at physical `grand unification' of
this wealth of observational phenomena, and comment on possibilities for
{\it Chandra}'s next decade in this field.
\end{abstract}






\section{Introduction}

The {\it Chandra} era has seen the proliferation of a greater variety
of possibly distinct observational classes of neutron star than ever
before (not even including accreting sources, ignored in this review).
With emission spanning the electromagnetic spectrum and radiative properties
that span a huge fraction of conceivable phase space,
such incredible range and diversity is not only unpredicted, but in 
many ways astonishing given the perhaps naively simple nature of the
neutron star.  Collapsed and close cousins to black holes,
why should neutron stars exhibit so much `hair'?  Indeed,
the sheer number of different class names is confusing even within the
field.  We therefore begin by presenting an introductory census
of the currently identified neutron zoo.

The classic young {\bf `radio pulsars'} (PSRs) pulse regularly and
predictably across the EM spectrum though generally are most observable
in the radio band.  The PSR prototype is often identified as the Crab
pulsar, even though it is far from prototypical of the class, given its
unparalleled energy output ($10^{38}$ erg/s),
and its magnificent nebula (see Figure 1).  The `radio pulsar'
term is not observationally accurate for the class, as, for example 
the `Geminga'
pulsar has all the properties of a radio pulsar except 
observable radio emission, possibly but not certainly due to
unfortunate beaming geometry \cite{hh92}.  The term {\bf `rotation-powered pulsar'}
(RPP) is therefore more cautious and precise: these objects are powered
by their loss of rotational energy due to braking by their magnetic fields. 

{\bf `Millisecond pulsars'} (MSPs), though also
rotation-powered, have different evolutionary histories, 
involving long-lived binary systems and a `recycling' accretion episode which spun-up
the neutron star and quenched its magnetic field \cite{bv91}.  In 
this review, MSPs are a sub-class of RPPs and have similar
emission properties, although
{\it Fermi} has recently shown that MSPs are surprisingly
bright $\gamma$-ray sources, suggesting refinements to
high-energy emission models are required \cite{aaa+09}.

The very recently discovered {\bf `rotating radio transients'} (RRATs)
\cite{mll+06} do not seem to produce observable periodic emission, the
defining property of RPPs.  Rather, they produce unpredictable sudden short radio 
bursts, which occur at integral multiples of an underlying periodicity.
These may be a sub-class of RPPs; this is discussed further below.

Radio emission was long thought to be
the hallmark of the non-accreting neutron star, Geminga notwithstanding.  
But several different sub-classes of generally radio-quiet neutron star have emerged 
in the {\it Chandra} era.  The {\bf `isolated neutron stars'} (INS; 
poorly named as most RPPs are also isolated but are not `INSs') have as
defining properties quasi-thermal X-ray emission with relatively low X-ray luminosity, great
proximity, lack of radio counterpart, and relatively long periodicities ($P=$3--11~s).  

Then there are the ``drama queens''
of the neutron-star population:  the {\bf `magnetars'}.  Magnetars have as their 
true defining properties occasional huge outbursts
of X-rays and soft-gamma rays, as well as luminosities in quiescence that are generally orders of magnitude greater
than their spin-down luminosities.  Magnetars are thought
to be young, isolated neutron stars powered ultimately by the decay of a very large magnetic field.
At their ``baddest,'' magnetars can briefly outshine all other cosmic soft-gamma-ray sources
{\it combined} \cite{hbs+05}.

The above census would be incomplete without mention of the handful of
X-ray bright {\bf `compact central objects'} (CCOs), a likely heterogeneous class.
They are so-named because of their central location in supernova remnants (SNRs), 
and have otherwise, until very recently, baffling properties.

The challenge of the past decade was -- and continues to be -- to find a way to
unify this variety into a coherent physical picture.
What determines whether a neutron star will be born with, for example, magnetar-like properties or
as a Crab-like pulsar?  What are the branching ratios for the various varieties, and,
given estimates of their lifetimes, how many of each are there in the Galaxy?
Ultimately such questions are fundamental to understanding the fate of massive stars,
and the nature of core-collapse, while simultaneously relating to a wider variety
of interesting high-energy astrophysics, ranging from the equation-of-state of
ultradense matter, to the physics of matter in ultrahigh magnetic fields.
In this contribution, we review the above observationally defined classes, and
consider the status of the progress toward `grand unification' of neutron stars into
a coherent physical understanding of the births and evolution of these objects.
Emerging as central to the picture are the stellar surface dipolar magnetic fields
(estimated via $B=3.2 \times 10^{19} (P \dot{P})^{1/2}$~G, where $\dot{P} = dP/dt$) 
and ages (estimated via the `characteristic age' $\tau \equiv P/2\dot{P}$).
We show how
observations at X-ray energies are essential to this work, and highlight, as much
as space permits, some of the great contributions made by {\it Chandra} this past decade.

\section{Rotation-Powered Pulsars (RPPs)}

RPP rotation periods span the range of 1~ms through 8~s.  Their magnetic
field strengths range from $10^8$~G to $9\times 10^{13}$~G.  Their distribution
in $P$-$\dot{P}$ space is shown in Figure 2.
Those with $P \lapp 20$~ms and $B \lapp 10^{10}$~G are usually called
MSPs.  The X-ray emission from RPPs falls into two broad
classes, both of which are generally pulsed and steady
(see \cite{krh06} for a review). 
First is thermal emission, which itself can be a result
of residual cooling following formation in a core-collapse
supernova (typically only observable for the first $\sim 10^5$~yr), 
or from surface reheating by return currents from
the magnetosphere, common for MSPs.  Thermal emission has received considerable attention owing to its
potential for use in constraining the equation-of-state of dense matter, by comparing
temperatures and luminosities with theoretical cooling curves, accounting for
the spectrally distorting impact of the neutron-star atmosphere, and also by
detailed modelling of X-ray light curves (e.g. \cite{bgr08}).
For reviews see, for
example, \cite{yp04} or \cite{zav07}.  {\it Chandra} highlights in this area includes
the detection of a surprisingly low temperature for the pulsar in the young
SNR 3C~58 \cite{shm02}, interesting constraints on the equation-of-state from
modelling millisecond pulsar thermal emission lightcurves 
\cite{bgr08}.  Major open questions in this area are the nature and impact of
the neutron-star atmosphere, and whether thermal-emission observations can 
strongly constrain the equation-of-state of dense matter.

Second is non-thermal, usually power-law emission originating
from the magnetosphere, typically more highly pulsed than the
thermal component, and strongly correlated with the pulsar's
spin-down luminosity.  The latter is defined as 
$\dot{E} \equiv 4 \pi^2 I \dot{P}/P^3$ where
$I$ is the stellar moment of inertia.  For a review of magnetospheric
emission, see \cite{har07}, although thinking on this front is
currently evolving thanks to recent interesting results
from {\it Fermi} \cite{aaa+09}.  {\it Chandra}'s primary strength in
this field is its small background and ability to resolve the point source from its nebular surroundings (see below), 
allowing superior spectral studies.  Major open questions on this subject
include how the emission is generated, and how is it related to observed
non-thermal $\gamma$-rays and radio emission.

{\it Chandra's} great angular
resolution means it has been a superb tool for studying RPP surroundings, namely
`pulsar wind nebulae' (PWNe), the often spectacular result of the
confinement of the relativistic pulsar wind by its environment (see
\cite{gs06} for a review).  Among {\it Chandra}'s greatest
PWN legacies are the discovery of rapid time variability in
the Crab and Vela PWNe (see \cite{wkp+07} and references therein) as well
as of the surprisingly diverse morphologies these objects can have
(see Figure 1 for examples and the aforementioned reviews for more).
Detailed modelling of the geometries \cite{nr08} and emission
mechanisms \cite{buc08} constrain both the properties of the
pulsar wind and shock acceleration mechanisms.  Important open
questions here are the nature and composition of the pulsar wind,
specifically whether it consists of only $e^+/e^-$ pairs or also
ions (e.g. \cite{aa06}), what are the particle energy distributions and what
is the fraction of energy in particles versus magnetic fields.
These have great relevance to our basic understanding of the neutron
star as a rapidly rotating magnetic dipole converting rotational kinetic
energy into such a powerful particle/field wind.
The shock acceleration issues are interesting as well (e.g. \cite{ss09}).

A subset of PWNe are the ram-pressure confined variety, showing bow-shock
morphology that corresponds with the pulsar's direction of motion through
the interstellar medium, e.g. \cite{kggl01,gvc+04}.  An example is shown in Figure 1. These are additionally
useful as they provide independent determinations of directions of motion,
which are often difficult to determine otherwise, especially for young pulsars.
Notably, the first-known PWN around an MSP, a ram-pressure confined structure
with morphology in clear agreement with the measured proper motion, was
found by {\it Chandra},
demonstrating unambiguously for the first time that MSPs
have winds that are similar to those of their slower cousins \cite{sgk+03}.

From a grand-unification perspective, the properties of RPPs can be seen
as the template against which all other classes are compared.  Thermal
emission, for example, is seen from all the other classes discussed
below, demonstrating that it is generic to neutron stars, except
the very cool ones for which it is unobservable.  By contrast, RPP-type
magnetospheric emission is only observable in high spin-down luminosity
sources, as are PWNe.  When considering any new neutron-star class, in
the absence of tell-tale pulsations, the spectrum is therefore crucial
to consider (thermal? non-thermal?) as is the presence or absence of
associated nebulosity.  This will be a recurring theme in the rest of
this paper.

\section{Rotating Radio Transients (RRATs)}

Recently, McLaughlin et al. \cite{mll+06} 
discovered brief radio bursts from Galactic sources,
but with no directly observed radio periodicities.  The number of `RRATs' is now roughly
a dozen, and in practically all cases, an underlying periodicity can be
deduced thanks to the phase constancy of the bursts.  At first thought
to be possibly a truly new class of neutron star, it now appears most
likely that RRATs are just an extreme form of RPP, which have long
been recognized as exhibiting sometimes very strong modulation of their
radio pulses \cite{wsrw06}.  Indeed several RRATs sit in unremarkable
regions of the $P$-$\dot{P}$ diagram (Figure 2).
Interesting though is the mild evidence for
higher-than-average periods and $B$ fields among the RRATs than in the
general population, though the numbers are small (Figure 2).  In {\it Chandra}
observations of the relatively high-$B$ RRAT J1819$-$1458 ($B=5\times
10^{13}$~G), the source had a thermal X-ray
spectrum typical of RPPs of the same age, arguing against any radical
physical difference \cite{rbg+06}.  Recently, 
a small apparently associated PWN 
was seen in {\it Chandra} data.  It has a somewhat surprisingly high
luminosity (though subject to substantial uncertainties), suggesting the presence of a possible
additional source of energy beyond rotation power \cite{rmg+09}.
If correct, it would be argue for RRATs being somehow physically
distinct from RPPs.

Regardless of whether or not RRATs are substantially physically different
from RPPs, their discovery is important as it suggests a very large population
of neutron stars that were previously missed by radio surveys which looked only
for periodicities.  This potentially has major implications for the neutron-star
birth rate \cite{mll+06,kk08}.

\section{Magnetars}

Magnetars, which traditionally are seen as falling into two classes, the
`anomalous X-ray pulsars' (AXPs) and the `soft gamma repeaters' (SGRs), have been studied 
with practically every modern X-ray telescope,
and {\it Chandra} is no exception.  
Today these long-$P$ objects are thought to be powered by an intense magnetic field
\cite{td95,td96a}, inferred
via spin-down to be in the range $10^{14}$-$10^{15}$~G (see Figure 2), well
above the so-called `quantum critical field' $B_{QED} \equiv m^{2}_{e} c^3/\hbar e = 4.4 \times 10^{13}$~G.
For a review of magnetars, see \cite{wt06}.  Magnetars have been argued to represent at most
$\sim$10\% of the neutron-star population, in part because of their apparent
association with very massive progenitors, as very strongly demonstrated
by the {\it Chandra} discovery of an AXP in the massive star cluster
Westerlund 1 \cite{mcc+06}.  However the discovery of the transient AXP
XTE~J1810$-$197 \cite{ims+04} suggests there could be a large, unseen population
of quiescent magnetars, which would have important implications for understanding
the formation of different types of neutron stars.

{\it Chandra}'s strengths for magnetar
observations have been its spectral resolution, allowing searches for spectral features
(none has been found; see \cite{pkw+01,jmcs02}), and its angular resolution for searching
for associated nebulae (none has been seen, with one possible exception; see \cite{vb09}),
and in some cases searching for a proper motion (none has been detected; see \cite{dceh09,kch+09}).  
More fruitfully, {\it Chandra} has provided precise
localizations for magnetars, crucial for multi-wavelength follow-up (e.g. \cite{ims+04,wpk+04}).
In addition, {\it Chandra}'s superior viewing windows relative to {\it XMM} have allowed rapid
target-of-opportunity (ToO) observations in response to magnetar outbursts, as triggered,
for example, by {\it RXTE} monitoring.  The latter telescope, with its excellent scheduling agility,
is ideal for regular snap-shot monitoring of magnetars, especially the generally bright
Anomalous X-ray Pulsars (AXPs) \cite{kcs99,kgc+01,gk02,dkg07,dkg08}.  However, {\it RXTE} provides only
pulsed fluxes above 2~keV, and spectroscopy of very limited quality in these short observations, hence
the need for occasional high-sensitivity focussing telescope observations.  For example, 
{\it Chandra} ToO observations
of AXP 1E 1048.1$-$5937 revealed a clear correlation between pulsed fraction and flux as well
as hardness and flux (see also \cite{tmt+05}) during the relaxation following this source's 2007 outburst.
Such studies constrain models of magnetar outbursts, generally confirming the ``twisted magnetosphere''
picture \cite{tlk02,tb05}.  Understanding the physics of magnetar outbursts is important
not just for studying the behavior of matter in quantum critical magnetic fields, but
also for constraining the burst recurrence rate, which is important for constraining
the number of magnetars in the Galaxy.  {\it Chandra}'s excellent point-source sensitivity has also been important
for studying transient magnetars in quiescence \cite{tkgg06}, showing that they can have
a huge dynamic range in flux and be very faint in quiescence, which further emphasizes the possibly large
unseen population of quiescent magnetars, important for determining
the magnetar birth rate (e.g. \cite{kk08}).  

\section{High-B Rotation-Powered Pulsars}

A potentially pivotal class of objects are the high-$B$ RPPs.  There
are now 7 RPPs (including one RRAT, J1819$-$1458) that have
spin-down inferred $B>4 \times 10^{13}$~G, a somewhat arbitrary limit
but comparable to $B_{QED}$, and very close
to the lowest $B$ field yet seen in a {\it bona fide} magnetar
($6\times 10^{13}$~G in AXP 1E~2259+586).  There is clear overlap between
high-$B$ RPPs and magnetars as is evident in the $P-\dot{P}$ diagram (Figure 2).
There have been multiple efforts to look for magnetar-like emission from
high-$B$ RPPs \cite{pkc00,msk+03,gklp04,km05,gkc+05,zkgl09}.  The
bottom line thus far is that in spite
of very similar spin parameters, the high-B RPPs show X-ray properties that are
approximately consistent within uncertainties with those of lower-$B$ RPPs of comparable age
(but see \cite{gkc+05}), though also consistent with spectra of transient magnetars
in quiescence \cite{km05}.  The latter point suggested that high-$B$ RPPs could
be quiescent magnetars.  The radio-detections of two transient magnetars lend
some support to this idea, although in those cases the radio emission appears temporarily
following an outburst, and the radio spectra are distinctly harder than those of most RPPs
\cite{crh+06,crhr07}

Recently, a major breakthrough occured.  The well
established young, high-$B$ ($B=4\times 10^{13}$~G) RPP PSR~J1846$-$0258 at the
center of the SNR Kes 75 exhibited
a sudden, few-week X-ray outburst in 2006, as detected by {\it RXTE} monitoring
\cite{ggg+08}.  The outburst included a large pulsed flux enhancement, magnetar-like
X-ray bursts \cite{ggg+08}, and, intriguingly, a coincidental unusual spin-up
glitch \cite{lkg10}.
This source had been monitored regularly with {\it RXTE} since 1999
\cite{lkgk06} and had shown no other activity.  {\it Chandra} had observed this
source twice: once in 2000 when the RPP was in its usual state, and by amazing
coincidence, once in 2006, in the few-week period of activity.  These {\it Chandra}
observations showed that the RPP X-ray spectrum softened significantly, with a thermal
component emerging from a previously purely power-law spectrum, and that the associated
PWN showed likely changes too, though possibly unassociated with the event
\cite{ggg+08,ks08,nsgh08}.
Overall, this event
demonstrates unambiguously a connection between high-$B$ RPPs and magnetars, further
supporting the possibility that high-B RPPs could in general be quiescent magnetars,
as speculated in \cite{km05}.

\section{Isolated Neutron Stars (INSs)}

The seven confirmed INSs are characterized, as mentioned earlier, by quasi-thermal X-ray
spectra, relative proximity (distances $\lapp 500$ pc), lack of radio or other counterpart, and 
relatively long spin-periods (3--11~s).  For past reviews of INSs, see \cite{hab07,kv09}. 
Recently a possible new example of INS was identified, in part thanks to {\it Chandra}
\cite{rfs08,sfr09}. These objects
are potentially very interesting for constraining the unknown equation-of-state (EOS) of
dense matter, as proper modelling of their X-ray spectra, which are thought to be
uncontaminated by magnetospheric processes, could constrain the stellar
masses and radii for comparison with different EOS models.  Also, INSs may
represent an interestingly large fraction of all Galactic neutron stars \cite{kk08}.
Timing observations of several objects, done in part by {\it Chandra}, have revealed that they are 
spinning down regularly, with inferred dipolar surface magnetic fields of typically a 
$\sim 1-3 \times 10^{13}$~G \cite{kv05a,kv09}, and characteristic ages of
$\sim$1--4 Myr (see Figure 2).  Such fields are somewhat higher than the typical
RPP field.  This raises the interesting question of why the closest neutron stars should have
preferentially higher $B$-fields.

The favored explanation for INS properties is that they are 
actually RPPs viewed well off from the radio beam.  Their X-ray luminosities are thought to
be from initial cooling and they are much less luminous than younger thermally cooling RPPs because
of their much larger ages.  However, their luminosities are too large for conventional
cooling, which suggests an addition source of heating, such as magnetic field decay, which
may explain their surprisingly high magnetic fields (discussed further below).

Particularly noteworthy in the INSs are puzzling broad and occasionally time-variable absorption features in their X-ray spectra
\cite{hztb04,zct+05,vkpm07,hnh+09}.  The {\it Chandra} LETG has contributed significantly
to this work (which is also being done using the {\it XMM} RGS).
The features have been suggested to be proton cyclotron lines, neutral hydrogen transitions,
but the time variability is puzzling, suggesting precession \cite{htd+06} or accretion episodes \cite{vkpm07}.
This remains an open issue.

\section{Central Compact Objects (CCOs)}

CCOs are neutron-star-like objects at the centers of supernova remnants,
but which have puzzling properties which have at some point precluded them from
being classified among one of the classes discussed above.  Common properties
to CCOs are absence both of associated nebulae and of counterparts at other
wavelengths.

The poster-child CCO, discovered in the {\it Chandra} first-light observation, 
is the central object in the young oxygen-rich
supernova remnant Cas A.  This mysterious object has been
very well studied, especially with {\it Chandra}.  Particularly
puzzling is its lack of X-ray periodicity, lack of associated nebulosity, 
and unusual X-ray spectrum
\cite{pza+00,cph+01,mti02,pl09}.  Recently, Ho \& Heinke \cite{hh09}
suggested that the Cas A CCO has a carbon atmosphere, showing
that its {\it Chandra}-observed spectrum is fit well by such a model,
and implies a stellar radius consistent with expectations for a neutron
star, in contrast to hydrogren or helium atmosphere models.  With the
entire surface radiating, the absence of pulsations would not be surprising.

Three other previously mysterious sources classified as CCOs are today,
thanks largely to {\it Chandra}, known to be pulsars with
interesting properties.  PSR J1852+0040 is at the center of
the SNR Kes 79 \cite{ghs05,hgcs07}.  This pulsar, observed only in
X-rays, has period 105~ms yet very small spin-down-inferred magnetic
field strength, $B = 3.1 \times 10^{10}$~G, and large characteristic age,
$\tau=192$~Myr \cite{hg10}.  This age is many orders of magnitude larger
than the SNR age, and much older than would be expected for an object
of this X-ray luminosity (which greatly exceeds the spin-down luminosity).
The $B$ value is also the smallest yet seen in young neutron stars,
prompting Halpern \& Gotthelf to call this an ``anti-magnetar.''
Interestingly, the object sits in a sparsely populated region of the $P$-$\dot{P}$ diagram
(Figure 2), among mostly recycled binary pulsars.
A similar case is the CCO in the SNR
PKS 1209$-$52, 1E 1207.4$-$5209.  Discovered in {\it ROSAT} observations
\cite{hb84}, pulsations were detected at 0.4~s in {\it Chandra} data
\cite{zpst00,pzst02}. 
Spectral absorption features were seen using {\it Chandra} \cite{spzt02}
and {\it XMM} \cite{dmc+04} but as of yet are not
well explained.  Interestingly, the source has an
unusually small (as yet undetectable period derivative), implying $B<3.3
\times 10^{11}$~G and $\tau_c>27$~Myr, again, orders of magnitude greater
than the SNR age and inconsistent with so large an X-ray luminosity \cite{gh07}.
Yet another such low-$B$ CCO is RX J0822$-$4300 in Puppis A \cite{gh09}.
Although its properties are not as well constrained ($B<9.8 \times
10^{11}$~G), it seems likely to be of similar ilk.  Gotthelf \& Halpern \cite{hg10} 
present a synopsis of other sources they classify as CCOs and argue that these
``anti-magnetars'' are X-ray bright thanks to residual thermal cooling
following formation, with the neutron star having been born spinning
slowly (consistent with some recent population synthesis studies,
e.g. \cite{fk06}) and with low $B$.  In this case, the origin of the
non-uniformity of the surface thermal emission remains puzzling.

Arguably the most bizarre CCO is 1E~161348$-$5055 in SNR RCW~103.
Discovered with {\it Einstein} \cite{tg80}, this CCO
showed no pulsations or a counterpart at other
wavelengths.  Unusually large variability was reported a decade
ago \cite{gpv99} and more recently,
a strong 6.6-hr periodicity was seen
from the source in an {\it XMM} observation \cite{dcm+06b}.  No infrared counterpart
has been detected in spite of deep observations enabled by a precise
{\it Chandra} position \cite{dmz+08}.  The nature of this source
is currently unknown, with suggestions ranging from an unusual binary
\cite{pcd+08,bg09} to a neutron star and a fall-back
disk \cite{li07}.


\section{Attempts at Unification}

In spite of the obviously great diversity in young neutron star observational
X-ray properties, some interesting ideas for grand unification are emerging.
A theory of magneto-thermal evolution in neutron stars has recently been developed
in a series of papers \cite{plmg07,apm08a,apm08b,pmg09}.  
Motivated largely by an apparent correlation between inferred $B$ field and
surface temperature in a wide range
of NS, including RPPs, INSs and magnetars \cite{plmg07} 
(but see \cite{zkgl09}), a model has recently been developed in which
thermal evolution and magnetic field decay are
inseparable.  Temperature affects crustal electrical resistivity, which
in turn affects magnetic field evolution, while the decay of the field
can produce heat that then affects the temperature evolution.
In this model, neutron stars born with large magnetic fields ($>5\times 10^{13}$~G)
show significant field decay, which keeps them hotter longer.  The magnetars
are the highest $B$ sources in this picture, consistent with observationally
inferred fields;  the puzzling fact that INSs, in spite of their
great proximity, all appear to have high inferred $B$s relative to the RPP
population, is explained nicely as the highest $B$ sources 
remain hottest, hence most easily detected, longest.  A recent attempt at
population synthesis modelling of RPPs, INSs and
magnetars using the magneto-thermal evolutionary models \cite{ppm+10}
followed closely a previous analysis of RPPs \cite{fk06} but found that the populations
of RPPs, INSs and magnetars (though they admit the latter two classes are few 
in number) can be explained if the birth magnetic field distribution of
nascent neutron stars has mean $B$ of $10^{13.25}$~G.
This is significantly higher than has been previously thought \cite{fk06}.

If this unification picture is correct, then RPPs, INSs and magnetars can
be roughly understood as having such disparate properties simply because of their
different birth magnetic fields and their present ages.
Unifying further, RRATs are likely just an
extreme form of RPP, with MSPs being binary-recycled RPPs with quenched
$B$ fields, as has been believed for many years \cite{bv91}.

The low-$B$ CCOs could be understood in the above picture as being the
lowest birth $B$ neutron stars, X-ray bright only because of their true
young ages, which are much smaller than their characteristic ages \cite{hg10}.
This however presents an interesting quandary.  The Kes~79 pulsar
(see above) with its small $B=3.1 \times 10^{10}$~G spins down very
slowly.  It is widely believed that that RPPs, as they spin down,
eventually cross a ``death line'' beyond which their radio emission
shuts off, in keeping with the absence of pulsars at long $P$ and small
$\dot{P}$ in the $P-\dot{P}$ diagram (Fig. 2).  There have been many
theory-motivated suggestions for the exact location of the death line
which is highly constrained by observations; one possibility
is shown in Figure 2 and is given by $B/P^2 = 0.17 \times 10^{12}$~G
(see \cite{fk06} and references therein).  For the Kes~79 pulsar, in the absence of $B$-field
decay (reasonable for so low a field in the magneto-thermal evolutionary
model) and assuming dipolar spin-down, it will have 
$P \sim 0.43$~s at death, which will occur in $\sim$3~Gyr.  Assuming the
other CCOs have similar properties, and noting that their parent SNRs
all have ages $\lapp$7~kyr, we estimate a birthrate of such low-$B$
objects of $\sim$0.0004~yr$^{-1}$.   For this birthrate and estimated
lifetime, we expect to find $\gapp 1 \times 10^6$ of these objects
in the Galaxy!  This is comparable to, or greater than the expected
number of higher-$B$ sources, which though born more often, do not live
as long.  Yet -- and here is the quandary -- the Kes~79 pulsar sits in
a greatly {\it underpopulated} region of the $P-\dot{P}$ diagram (see
Fig. 2).  Selection effects cannot explain this lack of pulsars; only
preferentially low radio luminosities could.  The latter is consistent
with models in which radio luminosity depends on spin-down parameters,
as opposed to being only dependent on viewing geometry (see \cite{fk06}).

\section{The Next Chandra Decade}

There is no shortage of ideas for future tests of the above theory of neutron star `grand
unification,' and {\it Chandra} is likely to play a major role.  The high-$B$
RPPs are a cross-roads class.  X-ray observations of these objects, both deep to determine
precise spectra and fluxes for comparison with the predictions
of magneto-thermal evolutionary models, as well as shallow monitoring to permit 
searches for magnetar-like variations, could be helpful.  
ToO observations of high-$B$ RPPs at
glitch epochs could reveal generic associated magnetar-like behaviour, which
would be a major breakthrough in our understanding of glitches, and possibly
neutron-star structure \cite{dkg08}.
Continuing ToO observations of magnetars in outburst will continue to 
constrain detailed models of magnetars and help understand crucial
parameters like outburst recurrence rate, central to understanding how
many magnetars there are in the Galaxy, hence their birthrate.  Continued monitoring of INSs,
both to study their intriguing spectra and their variations, as well
as their timing properties, is also important.  Systematic timing and
spectral studies of newly discovered members of any of the classes
discussed in this review, particularly CCOs, is clearly very important for making progress,
given the paucity of sources in most of the classes.  
With so many possibilities, the future of this field is very bright.
I look forward with high hopes to reading the
next {\it Chandra} decade's neutron-star review.





\begin{acknowledgments}
I take this opportunity to thank the {\it Chandra} staff for
their dedication, tirelessness, and extreme competence.
I acknowledge use of the ATNF Pulsar Catalog, the McGill AXP/SGR Online
Catalog and thank Dr. C.-Y. (Stephen) Ng for comments on this paper.  
I am grateful for support from the Lorne Trottier Chair in Astrophysics and Cosmology, 
from a Canada Research Chair, as well as from NSERC, FQRNT and CIFAR.  
\end{acknowledgments}






\end{article}



\begin{figure}[t]
\centering
\includegraphics[width=0.45\textwidth]{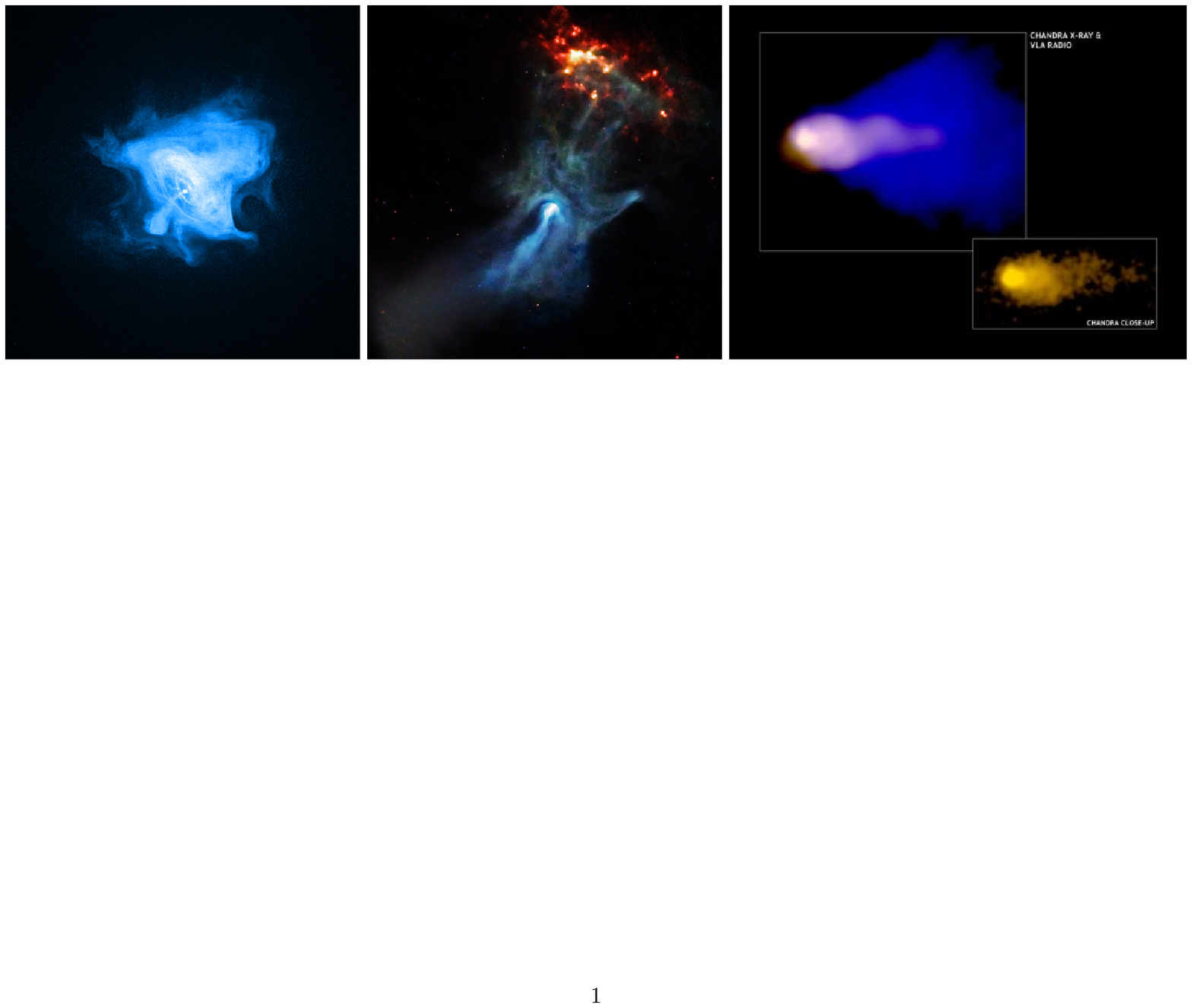}
\caption{Three examples of {\it Chandra}-observed pulsar wind nebulae.
(Left) The Crab Nebula (the image is 5$'$ across) with its clear
toroidal morphology and jet structure (NASA/CXC/SAO/F. Seward et al).
(Middle) The PSR B1509$-$58 pulsar wind nebula, knicknamed the ``hand of God'' (image
is 20$'$ across; NASA/CXC/SAO/P.Slane, et al., Ng et al. in prep.).
(Right) The `Mouse' ram-pressure-confined pulsar wind nebula
(1.2$'$ across; NASA/CXC/SAO/B.Gaensler et al. Radio: NSF/NRAO/VLA; \cite{gvc+04}). 
These images provide an indication of the variety of structures possible in PWNe.}
\end{figure}

\begin{figure}[h]
\centering
\includegraphics[width=0.45\textwidth]{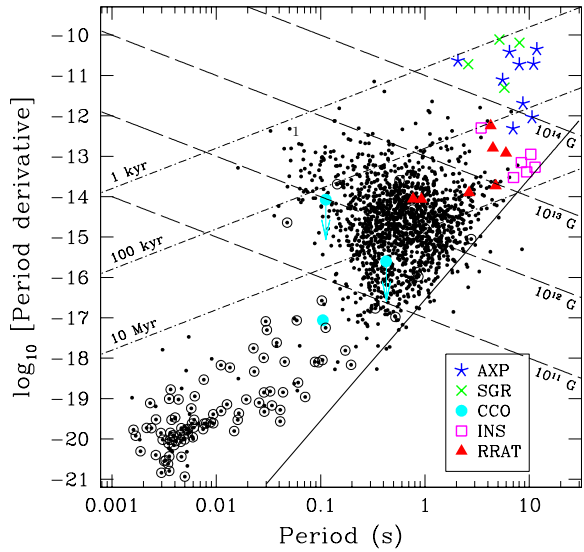}
\caption{$P$-$\dot{P}$ diagram for 1704 objects, including 1674 RPPs (small
black dots),
9 AXPs (blue crosses), 5 SGRs (green crosses), 3 CCOs (cyan circles), 6 INSs
(mageneta squares), and 7 RRATs (red triangles) for which these parameters
have been measured.  Open circles indicate binary systems.  Data from the ATNF Pulsar catalog
(www.atnf.csiro.au/research/pulsar/psrcat), the McGill SGR/AXP Online Catalog
(www.physics.mcgill.ca/~pulsar/magnetar/main.html), as well
as from \cite{hg10} and \cite{kv09}.  MSPs are RPPs having periods below
$\sim$20~ms.  Lines of constant $B$ (dashed) and $\tau$ (dot-dashed) are provided.
The solid line is a model death line (see text).}
\end{figure}






\end{document}